\documentclass{article}
\usepackage{spconf,amsmath,graphicx}
\usepackage{amssymb}
\usepackage{xcolor}
\usepackage{subcaption}
\usepackage{multirow}
\usepackage{caption}
\usepackage{url}

\usepackage{hyperref}
\hypersetup{colorlinks=true, linkcolor=black, citecolor=black, urlcolor=blue}
\title{A TRI-DYNAMIC PREPROCESSING FRAMEWORK FOR UGC VIDEO COMPRESSION }
\name{Fei Zhao\textsuperscript{1${\ast}$}\thanks{$^{\ast}$Intern in Bytedance. $^{\dag}$Corresponding author.}, Mengxi Guo\textsuperscript{2}, Shijie Zhao\textsuperscript{2${\dag}$}, Junlin Li\textsuperscript{2}, Li Zhang\textsuperscript{2}, Xiaodong Xie\textsuperscript{1}}
\address{Author Affiliation(s)}
\address{\textsuperscript{1}School of Computer Science, Peking University\\\textsuperscript{2}Bytedance
 }
\begin{document}
%\ninept
%
\maketitle
\begin{abstract}
In recent years, user generated content (UGC) has become the dominant force in internet traffic. However, UGC videos exhibit a higher degree of variability and diverse characteristics compared to traditional encoding test videos. This variance challenges the effectiveness of data-driven machine learning algorithms for optimizing encoding in the broader context of UGC scenarios.  
To address this issue, we propose a Tri-Dynamic Preprocessing framework for UGC.
Firstly, we employ an adaptive factor to regulate preprocessing intensity. Secondly, an adaptive quantization level is employed to fine-tune the codec simulator. Thirdly, we utilize an adaptive lambda tradeoff to adjust the rate-distortion loss function. Experimental results on large-scale test sets demonstrate that our method attains exceptional performance.
\end{abstract}
\vspace{-5pt}

\section{Introduction}
\label{sec:intro}
With the rapidly developing of social media and video-sharing platforms, user-generated content (UGC) has become a significant and expanding segment of internet traffic for video streaming services.
To tackle this problem, many researchers have contributed to the development of highly effective video coding algorithms. 
% Despite the development of video coding standards, such as H.264/Advanced Video Coding (AVC)~\cite{h264}, High-Efficiency Video Coding (HEVC)~\cite{HEVC}, and Versatile Video Coding (VVC)~\cite{VVC}, over the past few decades, the constantly growing amount of streaming content necessitates more effective video compression methods.
Despite some efforts made towards the development of video coding standards like H.264/AVC~\cite{h264}, H.265/HEVC~\cite{HEVC}, and H.266/VVC~\cite{VVC}, the ever-increasing volume of streaming contents demand more efficient video compression optimizations.

% Compared with optimization based on some traditional objective metrics, perceptual video coding is becoming one of the most promising research directions in video coding. This optimization is more in line with the perception of the human visual system (HVS). This research direction has not only given rise to a series of Video Quality Assessment (VQA) methods, such as Video Multi-method Assessment Fusion (VMAF)~\cite{VMAF}, but has also prompted the development of perceptual optimization algorithms for standard video codecs ~\cite{perceptual2, perceptual3, chadha2021deep}. In these research works, coding preprocessing is a worthwhile direction as it can improve coding efficiency without additional changes to the codec. 

Compared to traditional objective metrics, perceptual video coding is a promising research avenue aligning closely with the human visual system (HVS) perception. This systematic research not only includes Video Quality Assessment (VQA) methods like Video Multi-method Assessment Fusion (VMAF)~\cite{VMAF}, but also involves the development of perceptual optimization algorithms for standard video codecs~\cite{perceptual2, perceptual3, chadha2021deep}. Within these studies, preprocessing stands out as a valuable approach for enhancing coding efficiency without requiring codec modifications.

\begin{figure}
    \centering
    \includegraphics[width=8.5cm]{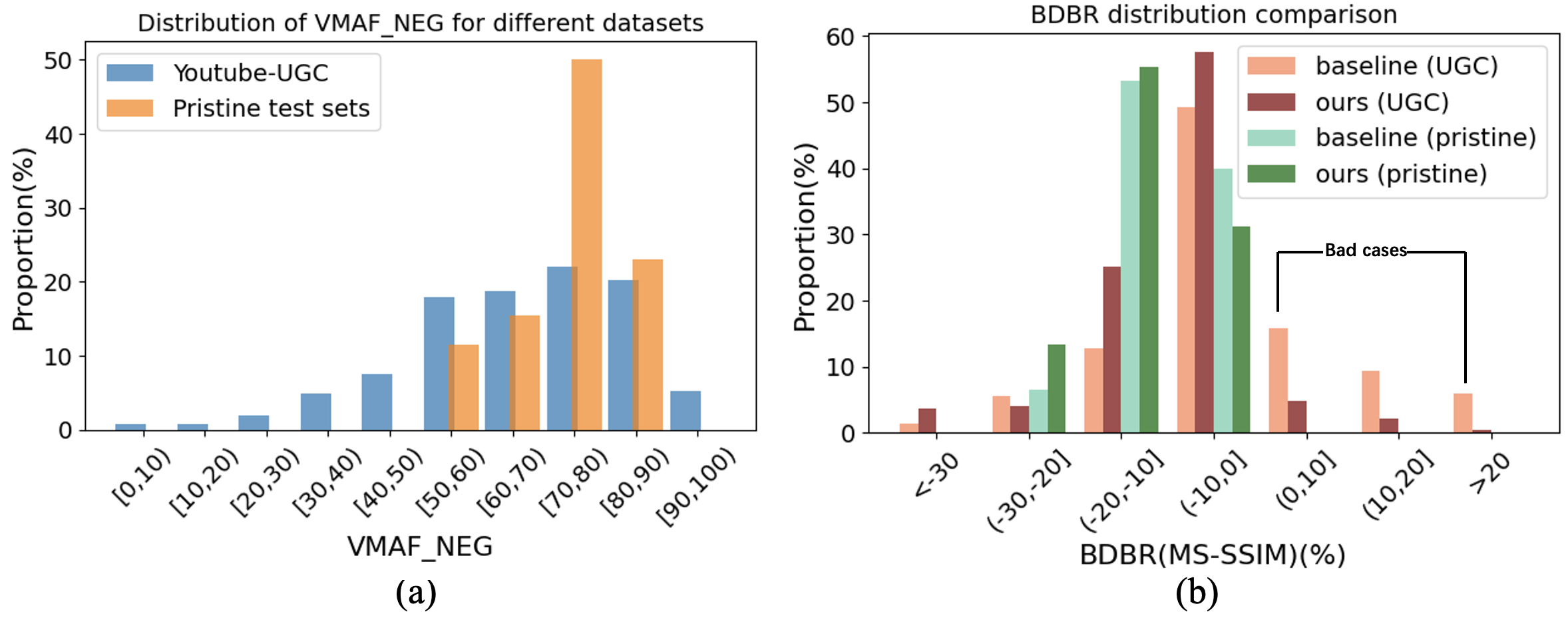}
    \captionsetup{skip=2pt}
    \caption{(a).Distribution of VMAF\textunderscore{}NEG after encoding different datasets under 1500kbps (b).The BDBR performance distribution of baseline method compared with our method.}
    \captionsetup[figure]{skip=-10pt}
    \label{fig:intro}
\end{figure}
% \vspace{-3pt}

With the advent of deep learning, preprocessing techniques utilizing deep neural networks have made strides in perceptual-rate optimization. In contrast to traditional preprocessing techniques such as Motion-Compensated Temporal Filtering (MCTF)~\cite{enhorn2020temporal}, deep preprocessing has demonstrated greater gains in Bjøntegaard Delta Bit Rate (BDBR)~\cite{bdrate} on pristine test sets~\cite{chadha2021deep}, in terms of perceptual metrics. 
It demonstrates the potential of deep learning in video preprocessing. However, issues arise when applying deep preprocessing to UGC data. As seen in Fig.~\ref{fig:intro}(a), the baseline model is the deep pre-processor which we reproduced based on~\cite{chadha2021deep}. While this model exhibits good performance on pristine test sets, it results in the occurrence of bad cases which means BDBR (MS-SSIM) $>$ 0 and accounted for up to 26\% of the total in the Youtube-UGC~\cite{wang2019youtube}. This phenomenon results from the fact that
% This illustrates the potential of deep learning in preprocessing. However, challenges emerge when applying deep preprocessing to UGC data. As depicted in Fig.~\ref{fig:intro}(b), the baseline model represents the deep pre-processor, which we reconstructed based on \cite{chadha2021deep}. While this model performs well on pristine codec testing sequences, it yields unsatisfactory results with BDBR (MS-SSIM) $>$ 0 for approximately 26\% of the videos in the Youtube-UGC dataset. The reason for this phenomenon is
UGC videos exhibit a higher spatio-temporal complexity compared to pristine test sets, leading to nonnegligible differences in encoding behaviour. As shown in Fig.~\ref{fig:intro}(b), we employ VMAF\_NEG as the distortion metric to compare the distribution of videos encoded at 1,000kbps on both the Youtube-UGC and pristine test sets (JVET Class B, XIPH~\cite{XIPHweb}, etc.). It is evident that UGC videos demonstrate greater diversity than pristine test sets. Consequently, previous deep preprocessing methods lack the ability to perceive the spatial and temporal information of video sequences, making them unsuitable for effectively handling the numerous UGC videos.

To address the aforementioned issue, this study proposes a Tri-Dynamic Preprocessing (TDP) framework for compressing UGC videos. The TDP framework comprises three components, namely Dynamic Processing Intensity (DPI), Dynamic Quantization Level (DQL), and Dynamic Lambda Trade-off (DlamT). These components exploit pre-analyzed features of the source contents in the training phase to perceive spatio-temporal complexity in video sequences, effectively training the deep preprocessing network. 
% During testing, only DPI is utilized to attain dynamic processing intensity control in deep preprocessing.
Experimental results demonstrate that our pipeline enables the preprocessing network to achieve 7.14\% BDBR reduction on VMAF\_NEG while 12.03\% BDBR reductions on VMAF. Moreover, our method can significantly reduce the number of bad cases on the YouTube UGC dataset.

\vspace{-5pt}

\section{Proposed Method}

\begin{figure}
    \centering
    \includegraphics[width=8.5cm]{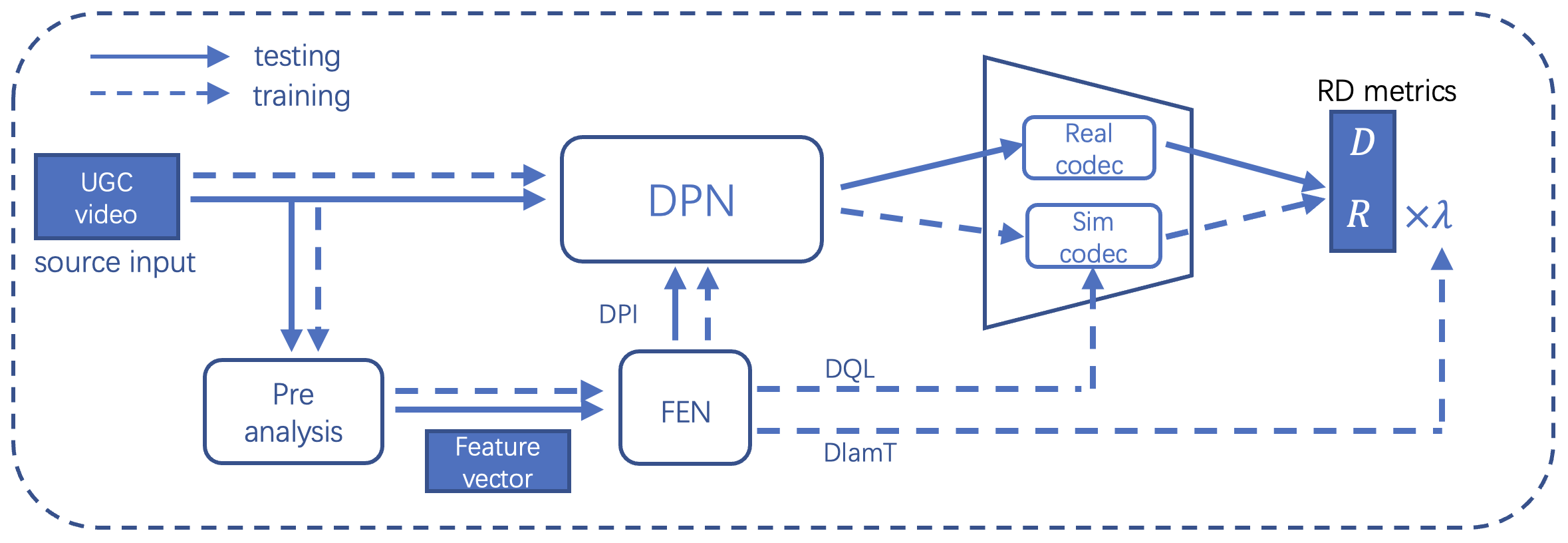}
    \captionsetup{skip=3pt}
    \caption{The overall pipeline of the proposed method. Note that we utilize the full TDP (DPI, DQL and DlamT) for training but only DPI while testing. We use a codec simulator for training and real codec for testing.}
    \label{fig:pipeline}
\end{figure}
As depicted in Fig.~\ref{fig:pipeline}, our TDP framework consists of pre-analysis and the three-fold dynamic scheme: DPI, DQL, and DlamT. All three TDP components are used in training, while only DPI is applied during testing. We will provide a detailed explanation of these components within the TDP in the following subsections.

\vspace{-9pt}
\subsection{Pre-analysis}

We perform pre-analysis on input videos to generate feature vectors, guiding the following tri-dynamic preprocessing during training. This involves extracting frame-level Spatial Information (SI) and Temporal Information (TI)~\cite{SITI}, including maximum, average, and standard deviation values. We also utilize x264~\cite{merritt2006x264} with preset \textit{fast} mode at a specified bitrate to obtain the frame-level $QP$.
% , motion vector bit usage (mv), intra-mode macroblock count (imb), inter-mode macroblock count (pmd), and skip mode macroblock count (smb). 
% The SITI features are condensed into DPI factors using a Feature Extraction Network (FEN) with 2-layer Multi-Layer Perceptrons (MLPs). The coding generated $QP$ will be further used to derive DQL and DlamT factors.
All features are concatenated together to form a 7-dimensional feature vector and then processed by a two-layer Multi-Layer Perceptron (MLP) Feature Extraction Network (FEN) to generate a dynamic intensity factor $f_{d}$ which is applied to the deep preprocessing network through DPI. Notably, the $QP$ feature undergoes further utilization in the DQL and DlamT.

% , resulting in DPI (Dynamic Processing Intensity) from SI/TI information and DQL (Dynamic Quantization Level) and DlamT (Dynamic Lambda Trade-off) from the 264 fast pass data. During testing, we exclusively employed the DPI factor derived from SI/TI information.

% We conducted a pre-analysis of the input videos to obtain a series of feature vectors, which serve as guidance for the TDP control factors during training. In this process, we extract frame-level SI/TI information from the input videos, including their maximum, mean, and standard deviation values. Additionally, we utilized the x264 \cite{merritt2006x264} fast encoding mode for a rapid one-pass encoding at a specified bitrate, resulting in valuable coding-oriented features. These features encompassed QP (Quantization Parameter), mv (motion vector bit usage), imb (intra mode macro block count), pmb (inter mode macro block count), and smb (skip mode macro block count).

% The features generated from this pre-analysis are then distilled into three factors using the feature extraction network (FEN) which contains shallow MLPs (Multi-Layer Perceptrons) with two hidden layers. Specifically, the SI/TI information gave rise to the DPI (Dynamic Processing Intensity) factor, while the 264 fast pass information yielded DQL (Dynamic Quantization Level) and DlamT (Dynamic Lambda Trade-off). It's worth noting that during testing, we solely utilized the DPI factor derived from the SI/TI information.
\vspace{-9pt}
\subsection{Dynamic Processing Intensity}

\begin{figure}
    \centering
    \includegraphics[width=8.5cm]{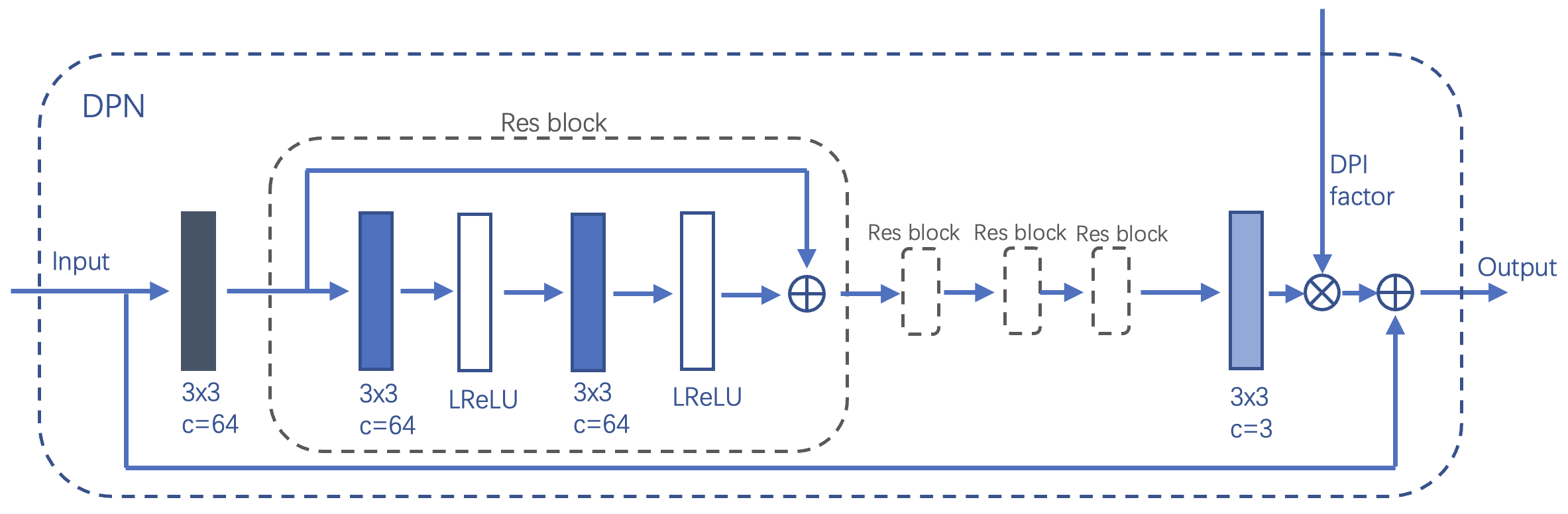}
    \captionsetup{skip=8pt}
    \caption{The structure of proposed DPN which contains four basic residual blocks. By employing the residual connection architecture, we can dynamically fine-tune the processing intensity through a multiplicative factor.}
    \label{fig:DPN}
\end{figure}

% A video encoder $\mathcal{C}$ is commonly perceived as a lossy compressor, implying that distortion $D$, exists between its output encoded video $\mathcal{C}(I)$ and its input source video $I$, as shown as $D=|I-\mathcal{C}(I)|$. In the context of preprocessing, when the source video undergoes processing via a preprocessor $\mathcal{P}$, resulting in a processed video $\mathcal{P}(I)$, the overall system distortion becomes $D_{pre+c}=|I-\mathcal{C}(\mathcal{P}(I))|$.

We model the preprocessing optimization procedure. When employing a rate-distortion (RD) oriented preprocessing $\mathcal{P}$ to a video $x$, if a target bitrate is specified, one of the primary objectives is to minimize the overall system distortion, expressed as $min(D_{proc+codec}=|x-\mathcal{C}(\mathcal{P}(x))|)$.
In certain specific scenarios, particularly when the spatio-temporal complexity of the input video sequence is exceptionally low, the encoder $\mathcal{C}$ operates in an almost lossless manner. 
During such instances, the distortion $D_{codec}$ approaches zero, thereby:
\vspace{-1pt}
\begin{equation}
    D_{codec}=|x-\mathcal{C}(x)|\rightarrow0; 
\end{equation}
\begin{equation}
   D_{proc+codec}\rightarrow|\mathcal{C}(x)-\mathcal{C}(\mathcal{P}(x))|.
\end{equation}

In such cases, $\mathcal{C}(x)\rightarrow\mathcal{C}(\mathcal{P}(x))$, indicating that $\mathcal{P}(x)$ approaches $x$. This implies that the preprocessing intensity should be reduced to zero. 
Employing a fixed preprocessing intensity, as seen in previous approaches, clearly lacks rationality. 
Here we propose a dynamic processing intensity (DPI) scheme to solve this issue. 
% It is imperative to devise a mechanism that allows the preprocessing intensity to autonomously adapt based on the characteristics of the input video. This adaptation aims to achieve superior overall performance.
We architect the network structure of the dynamic preprocessing network (DPN) in the form of residual connections, as shown in Fig.~\ref{fig:DPN}. 
% This design involves subjecting input frames to a sequence of convolutions, resulting in a feature map. This feature map is then combined with the original input through an addition operation to generate the processed frame. 
DPN is responsible for the preprocessing of the video sequence and outputs a residual processing mask $x_{m}$. To enable the preprocessing to regulate processing intensity, we manipulate the magnitudes of $x_{m}$ within the dynamic scaling factor from FEN as
\vspace{-1pt}
\begin{equation}
    \mathcal{P}(x)=f_{d}\times x_{m} + x.
\end{equation}

As deduced from our earlier analysis, this factor, synonymous with the processing intensity, exhibits a close correlation with the spatio-temporal characteristics of the input video. 
Building upon this foundation, we utilize the factor $f_{d}$ 
derived from features as mentioned in section 2.1, serving as a means to control the processing intensity.
% To this end, we utilize the spatial information (SI)\cite{SITI} and temporal information (TI)\cite{SITI} of the input video as indicators for determining the value of the multiplicative factor. To be more specific, we extract frame-level SI/TI information (mean value \& standard deviation value) and channel this through a two-layer MLP (Multi-Layer Perceptron) to yield a predictive scale factor value.
\vspace{-9pt}
\subsection{Dynamic Quantization Level}
% The second aspect of the adaptive control mechanism involves adjusting the quantization level within the codec simulator, a pivotal factor in supervising distortion during the preprocessor's training phase. 
The quantization factor holds a key position as a hyperparameter within the preprocessor's training process. A higher quantization level leads to a more pronounced quantization distortion perceived by the preprocessor, consequently resulting in a heightened processing intensity of the trained preprocessor, and vice versa.
In this context, the aim is to adapt the quantization factor according to the quality of the input video. 
% To be more precise, under a specified target bitrate, videos with lower complexity—easier to encode—tend to more readily meet the bitrate requirements. This, in turn, allows for a lower quantization intensity. Conversely, videos with higher complexity necessitate a more stringent quantization intensity. 
To achieve this, we propose a dynamic quantization level (DQL) scheme, which employs a dynamic quantization factor $f_{q}$ to adjust the quantization level of the codec simulator while training. 
% As mentioned in section 2.1,
% a rapid coding pass is performed at a fixed bitrate to extract coding-oriented features for FEN to generate the $f_{DQL}$. In particular, we limit the max value of $f_{DQL}$ to 50.
As mentioned in section 2.1,
a rapid coding pass to catch the frame-level $QP$ of input video at a specific bitrate, then the $QP$ value is clipped to range [1,50] as the value of $f_{q}$ which is applied to the codec simulator, s.t. $f_{q} = clip(QP)$.

%In particular, we limit the max value of $f_{DQL}$ to 50.
% \vspace{-10pt}

\vspace{-4pt}

\subsection{Dynamic Lambda Trade-off}
Our optimization loss function is the rate-distortion loss as:
\vspace{-3pt}
\begin{equation}
    L=D + \lambda \times R,
    % \vspace{0.09cm}
\end{equation}
\vspace{-1pt}
where $D$ is the error between source video $x$ and compressed pre-processed video $\mathcal{C}(\mathcal{P}(x))$, and $R$ is the bits used for encoding the pre-processed video. Notably, $\lambda$ holds crucial importance as it represents the trade-off between bit rate and distortion. Previous deep learning-based preprocessing or compression works~\cite{chadha2021deep, DVC} often use a fixed $\lambda$, which makes it difficult for the neural network to handle extreme cases. To address this issue, we propose a Dynamic lambda Trade-off (DlamT) strategy. Specifically, inspired by~\cite{HEVC, VVC}, we model the logarithmic result of $\lambda$ and $QP$ as a fixed positively-correlated linear mapping, i.e., $\log _{10}(\lambda_{adpt})=k \times QP+b$. We set $k=0.12$ and $b=-8$ based on our extensive experimental results. With this setting, we constrain the adaptive lambda $\lambda_{adpt}$ values within the range of (1e-8, 1e-2], corresponding to frame-level $QP$ with a range of (0, 50]. It should be noted that the range of $\lambda_{adpt}$ values depends on the values of ${D}$ and ${R}$ selected. In our experiments, we adopt MS-SSIM loss~\cite{MS-SSIM} as the measurement of distortion $D$ and use the learnable factorized prior entropy model of Balle et al.~\cite{balle2018variational,balle2016end} to calculate $R$. The final rate-distortion loss is written as:
\vspace{-2pt}
\begin{equation}
    L=D(\mathcal{C}(\mathcal{P}(x)), x) + \lambda_{adapt} \times R.
    % \vspace{0.09cm}
\end{equation}

\begin{table*}\small
\begin{center}
\caption{R-D performance (BDBR\%) on Youtube UGC dataset with VVC.}
\label{main}
% \begin{tabular}{| X | X | X | X | X | X | X | X | X | X | X | X | X |}
\begin{tabular}{| c || c | c | c | c || c | c | c | c || c | c | c | c |}
% \hline
%  & BD-Rate reduction & BD-Rate reduction\\
%  & vs. H.264 & vs. H.265 \\
\hline
 BDBR & \multicolumn{4}{|c||}{MCTF~\cite{enhorn2020temporal}} & \multicolumn{4}{|c||}{Baseline} & \multicolumn{4}{|c|}{Ours}\\

% & structure & size & (PSNR)($\%$) & (SSIM)($\%$)\\ 
\hline
\multirow{2}{*}{Category} & MS- & \multirow{2}{*}{SSIM} & VMAF\textunderscore{} & \multirow{2}{*}{VMAF} & MS- & \multirow{2}{*}{SSIM} & VMAF\textunderscore{} & \multirow{2}{*}{VMAF} & MS- & \multirow{2}{*}{SSIM} & VMAF\textunderscore{} & \multirow{2}{*}{VMAF}\\
& SSIM &  & NEG &  & SSIM &  & NEG &  & SSIM &  & NEG & \\
\hline

Animation & -2.57& -2.8 &  -2.12& -2.88 & -2.77& -2.29& -6.91&\textbf{-12.65} & \textbf{-4.11}& \textbf{-8.47}& \textbf{-8.24}& -12.56\\
\hline
CoverSong & -1.87 & -1.89 &  -1.69& -1.7 & -1.1& -2.31& -4.48&-7.41 & \textbf{-2.49}& \textbf{-5.99}& \textbf{-7.33}& \textbf{-10.24}\\
\hline
Gaming& -2.29 & -2.64 &  -2.33& -2.3 & -5.93& -8.07& -5.41& -11.52& \textbf{-9.17} & \textbf{-13.15} & \textbf{-10.58} & \textbf{-14.95} \\
\hline
HowTo& -0.47 & -0.48 &  -0.42& -0.35 & 8.05& 7.46& 6.76& -4.85& \textbf{-0.71}& \textbf{-3.57}& \textbf{-2.56} & \textbf{-10.85}\\
\hline
Lecture& \textbf{-4.7} & -4.04 &  \textbf{-5.0}& -6.67 & -2.16& -3.84& -0.97& \textbf{-12.21}& -3.41& \textbf{-13.9}& -4.15& -9.68\\
\hline
LiveMusic& -1.7 & -2.42 & -1.30 & -1.36 & 4.88& 10.69& -4.17& -6.74& \textbf{-6.42}& \textbf{-8.62}& \textbf{-9.46}& \textbf{-11.07}\\
\hline
LyricVideo& \textbf{-0.4} & -0.45 &  -0.41& -0.37 & 12.43 & 8.53 & -0.81& \textbf{-19.73}& -0.13 & \textbf{-1.24} & \textbf{-3.14}& -6.8\\
\hline
MusicVideo& -1.86 & -1.93 &  -1.39& -1.3 & -1.04 & -0.17 & -2.82 & -5.49 & \textbf{-4.64}& \textbf{-5.44}& \textbf{-5.08}& \textbf{-7.16}\\
\hline
NewsClip& -2.11 & -2.1 &  -2.06& -2.01 & -0.45 & -4.15 & -8.33 & -20.18 & \textbf{-3.22}& \textbf{-9.68}& \textbf{-11.62}& \textbf{-20.38}\\
\hline
VerticalVideo& -0.48 & -0.48 &  -0.34& -0.26 & 5.66 & 5.36 & -1.2& \textbf{-12.52} & \textbf{-0.66}& \textbf{-2.26}& \textbf{-3.72}& -5.85 \\
\hline
Sports& -1.3 & -1.32 &  -1.34& -1.29 & -3.21&-5.18&-6.3&-9.41& \textbf{-3.81}& \textbf{-8.24}& \textbf{-8.52}& \textbf{-10.27}\\
\hline
TelevisionClip& -1.22 & -1.21 &  -1.1& -0.95 & -0.33&-2.32&-6.5&-13.99& \textbf{-2.87}& \textbf{-6.75}& \textbf{-8.48}& \textbf{-14.89}\\
\hline
Vlog& -2.75 & -2.85 &  -1.96& -1.82 & -2.09&-1.32&-4.49&-7.0& \textbf{-4.38}& \textbf{-5.42}& \textbf{-6.0}& \textbf{-7.64}\\
\hline\hline
Avg& -1.88 & -1.95&  -1.75& -1.86 & -0.21 & -1.03 & -3.79 & -10.61 & \textbf{-4.07}& \textbf{-7.78}& \textbf{-7.14}& \textbf{-12.03}\\
% \hline
% Bad case ratio & 10.8 & 10.4 & 12.8 & 14.5 & 26.8 & 27 & 12.4 & 12 & 7.2 & 8.5 & 6.7 & 6.5\\
\hline
\end{tabular}
\captionsetup{skip=9pt}
\end{center}
\end{table*}

\vspace{-10pt}

\begin{table}
\begin{center}
\caption{R-D performance for H.264 and H.265.}
\label{more_codec}
\resizebox{0.46\textwidth}{!}{
\begin{tabular}{| c | c | c | c | c | c |}
% \hline
%  & BD-Rate reduction & BD-Rate reduction\\
%  & vs. H.264 & vs. H.265 \\
\hline
 Codec & Methods & MS-SSIM & SSIM & VMAF\textunderscore{}NEG & VMAF \\

% & structure & size & (PSNR)($\%$) & (SSIM)($\%$)\\ 
\hline
\multirow{2}{*}{H.264}  & Baseline & -0.76 & -2.8&  -6.03 & -11.5 \\
\cline{2-6}
 & Ours & -3.08 & -8.46 & -7.84 & -12.84 \\
\hline
\hline
\multirow{2}{*}{H.265} & Baseline & -1.21 & -2.29& -5.8 & -9.64 \\ 
\cline{2-6}
 & Ours & -3.98 & -8.09 & -8.11 & -12.35\\
\hline

\end{tabular}}
\end{center}
\end{table}

\vspace{-1pt}

% \begin{table*}
% \begin{center}
% \caption{R-D performance for more codecs}
% \label{more_codec}
% \begin{tabular}{| c | c | c | c | c | c | c | c | c |}
% % \hline
% %  & BD-Rate reduction & BD-Rate reduction\\
% %  & vs. H.264 & vs. H.265 \\
% \hline
%  BDBR & \multicolumn{4}{|c|}{Baseline} & \multicolumn{4}{|c|}{Ours}\\

% % & structure & size & (PSNR)($\%$) & (SSIM)($\%$)\\ 
% \hline
% Codec & MS-SSIM & SSIM & VMAF\textunderscore{}NEG & VMAF & MS-SSIM & SSIM & VMAF\textunderscore{}NEG & VMAF \\
% \hline

% H.264 & -0.76 & -2.8&  -6.03 & -11.5 & -3.08 & -8.46 & -7.84 & -12.84\\
% \hline
% H.265 & -1.21 & -2.29& -5.8 & -9.64 & -3.98 & -8.09 & -8.11 & -12.35\\
% \hline

% \end{tabular}
% \end{center}
% \end{table*}
\vspace{-1pt}
\section{Experiments}
\vspace{-1pt}
\subsection{Implementation}
% To better represent the UGC domain in our training data, we uniformly sample over 6000+ video clips from the largest publicly available academic UGC video dataset, YouTube-8M \cite{abu2016youtube}. From these video sequences, we generate a training dataset comprising 20,000+ sub-sequences, each consisting of three frames. 
To better represent the UGC domain in our training data, we generate a training dataset from Youtube-8M~\cite{abu2016youtube} comprising 20,000+ sub-sequences, each consisting of three frames. 
Regarding the test dataset, we employ the most representative publicly available academic UGC dataset YouTube-UGC. For our testing purposes, we utilize 266 videos with 13 categories from the 1080p subset of YouTube-UGC.
% The whole categories include Animation, Cover Song, Gaming, How-to, Lecture, Live Music, Lyric Video, Music Video, News Clip, Sports, Television Clip, Vertical Video, and Vlog. 
% Each video category has compression-focused features. For instance, Gaming videos exhibit relatively higher motion, while Lyric and How-to videos often have static backgrounds. 
% This indicates that 
The dataset encompasses a sophisticated spatio-temporal complexity distribution~\cite{wang2019youtube}, qualifying it as a representative sample of UGC scenarios.

We construct our baseline with reference to~\cite{chadha2021deep} since there is no open-source deep preprocessing implementation available. For the codec simulator, we implement a version based on the descriptions in~\cite{chadha2021deep} and the framework provided in~\cite{balle2016end}. As seen in Fig.~\ref{fig:intro}(b), our baseline has achieved performance on the CTC dataset that is comparable to that reported in the paper. For the implementation of the pre-analysis, we use the Sobel filter to extract SI/TI feature~\cite{SITI} and employ x264 with a fast encoding preset with a constant bitrate of 1,500 kbps. The entire pipeline is implemented using the PyTorch deep learning framework and tested after training for 20 epochs with 4 NVIDIA A100-SXM-80GB GPUs.

In terms of testing, we replaced the codec simulator with standard codecs to perform RD results. For standard codecs, we use the open-source version of VVC (vvenc)~\cite{wieckowski2021vvenc}, H.265 (x265)~\cite{HEVC} and H.264 (x264)~\cite{h264} via preset \textit{medium} and employ the Constant Bit Rate (CBR) with widely used target bitrates of [1000, 2500, 4000, 5000] kbps. We use Bjøntegaard Delta Bit Rate (BDBR)~\cite{bdrate} as the method for measuring the RD performance, where the anchor is to compress videos with standard codecs without preprocessing.

\begin{figure}
    \centering
    \includegraphics[width=8.5cm]{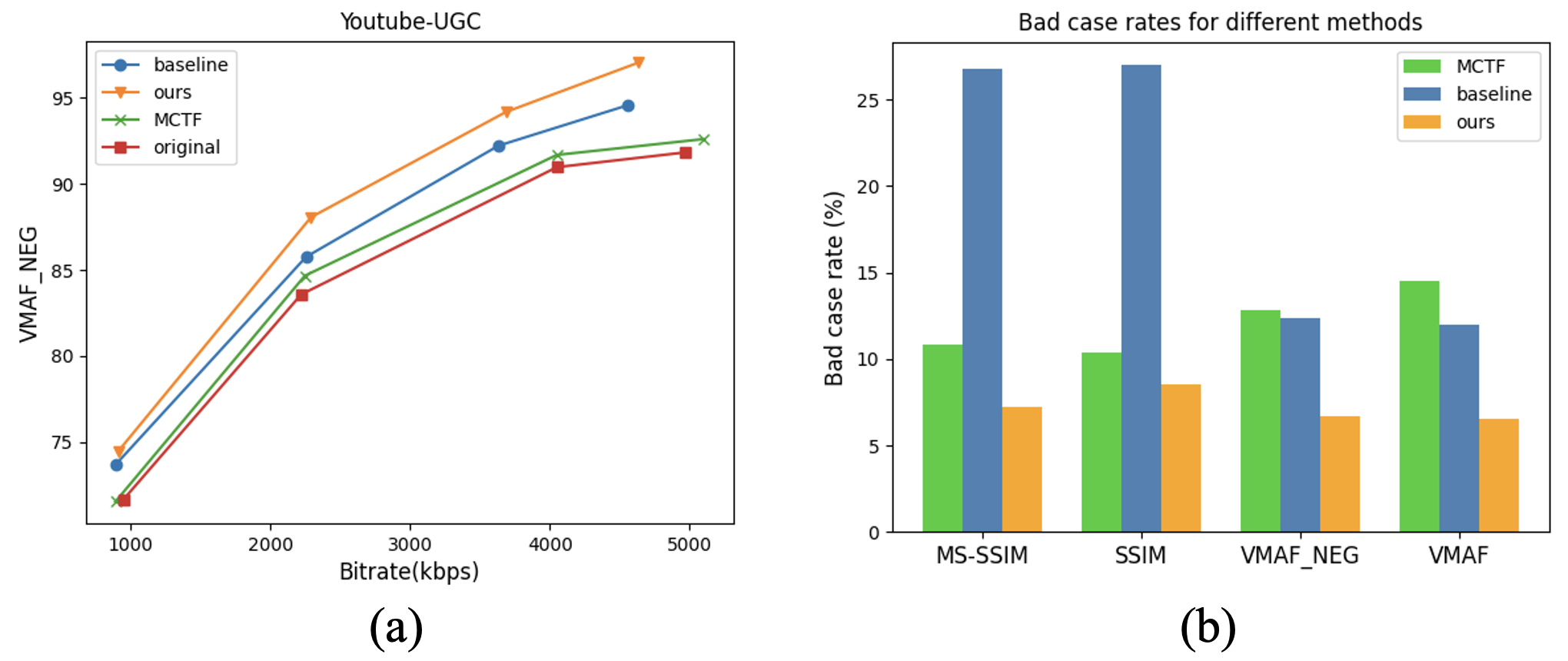}
    \captionsetup{skip=3pt}
    \caption{(a)Illustration of RD curves on VMAF\_NEG with VVC for Youtube-UGC;(b)Bad case rate for different methods with VVC, under Youtube-UGC dataset.}
    \captionsetup[figure]{skip=-3pt}
    \label{fig:badcase}
\end{figure}
\vspace{-7pt}
\subsection{Experimental Results}
To assess the overall performance of our proposed method, we conduct RD performance tests comparing our method with both the baseline method and the MCTF, an available internal tool of the VVC. All these methods are compared to encoding the source videos directly without any preprocessing or prefiltering.
Regarding the evaluation metrics, we employ four widely used perceptual metrics: MS-SSIM, SSIM, VMAF\textunderscore{}NEG, and VMAF. 
The RD results are presented in Table~\ref{main}, our method largely surpasses others in overall performance. Correspondingly, an illustration of RD curves for VMAF\textunderscore{}NEG can be seen in Fig.~\ref{fig:badcase}(a).
Our proposed method exhibits remarkable improvement in MS-SSIM and SSIM metrics in categories with low spatio-temporal complexity such as LyricVideo and How-to. We attribute this to the proposed TDP effectively perceiving the unique spatio-temporal characteristics of such videos. And it makes optimal adjustments, resulting in a substantial performance gain over the baseline. Notably, the gain of our method in terms of VMAF is not as good as the baseline in some categories. This can be attributed to the slight hack of the preprocessing algorithm on the VMAF metric~\cite{VMAFHacking}. The performance of our method on other metrics provides stronger evidence of our superiority.

We compute the proportion of bad cases (BDBR $>$ 0) for each evaluation metric. As shown in Fig.~\ref{fig:badcase}(b), our method substantially outperforms the MCTF and baseline method, accompanied by a notable reduction in the proportion of bad cases.
We also present the superior performance results under earlier standard codecs, H.264(x264) and H.265(x265)~\cite{hu2014analysis}, in Table~\ref{more_codec}. 

Additionally, we conduct tests to evaluate the individual impact of each part within the proposed TDP mechanism. The results, as shown in Table~\ref{ablation}, reveal that while each part can yield certain benefits independently, it is evident that their combined effect maximizes their potential performance.

\begin{table}

\begin{center}
\caption{Ablation study for different control schemes.}
\label{ablation}
\resizebox{0.46\textwidth}{!}{
\begin{tabular}{|c|c|c|c|c|c|c|c|}
    \hline
     DPI & DQL & DlamT & MS-SSIM & SSIM & VMAF\textunderscore{}NEG & VMAF\\
    \hline
    \checkmark & $\times$ & $\times$ & -0.28& -1.77& -4.01&-10.25\\
    \hline
    $\times$ & $\checkmark$ & $\times$ & -1.26 & -2.45 & -4.55 & -10.39\\
    \hline
      $\times$  & $\times$ & $\checkmark$ & -0.97 & -2.89 & -5.64 & -10.2\\
    \hline
    $\checkmark$& $\checkmark$ & $\checkmark$ & -4.07 & -7.78 & -7.14 & -12.03\\
    \hline
    
\end{tabular}
\captionsetup{skip=-20pt}
}
\end{center}
\end{table}

\vspace{-10pt}

\begin{figure}
    \centering
    \includegraphics[width=6cm]{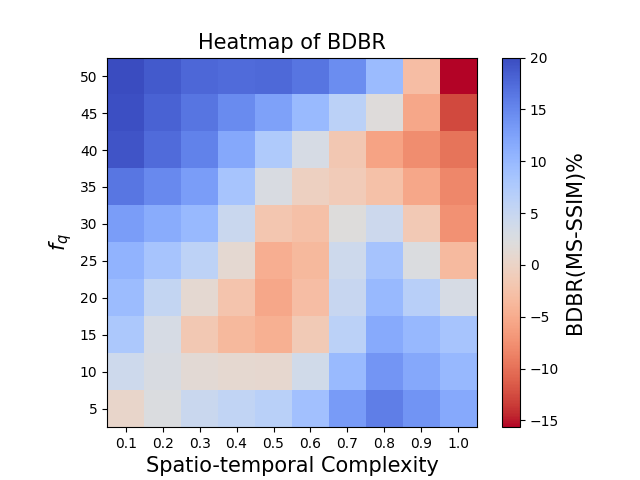}
    \captionsetup{skip=4pt}
    \caption{BD-Rate performance distribution heat-map for different spatio-temporal complexity and $f_q$.}
    \label{fig:heat}
\end{figure}

\vspace{-7pt}

\subsection{Further Discussion}

\vspace{-3pt}
% \begin{figure}
%     \centering
%     \includegraphics[width=7cm]{siti_aq.png}
%     \caption{Correlation of SITI-QP distribution.}
%     \label{fig:siti-aq}
% \end{figure}

% \vspace{-5pt}

% \vspace{-5pt}

% We conducted an analysis of the correspondence between SITI and QP distributions at a specific bitrate. As depicted in Fig.\ref{fig:siti-aq}, we gathered frame-level QP and SITI data by imposing a rate limit of 1500kbps on the Youtube-UGC test set. The results align with our previous observations, revealing a significant positive correlation between QP and SITI. In other words, images with larger SITI values tend to have larger QP values under the conditions of bitrate-constrained encoding. This validates the rationale behind our approach of training the preprocessor using the correspondence between SITI and QP.

We conduct additional experiments to analyze the relationship between source video characteristics, $f_{q}$, and the RD performance. 
We refer to the method in~\cite{wang2019youtube} for calculating coding-oriented spatio-temporal complexity.
%, which involves normalizing I-frame bits and normalized P-frame bits to obtain a fair indicator of spatial-temporal complexity.
We divide the normalized spatio-temporal complexity into ten groups with a width of 0.1. For each group of videos, we found the corresponding $f_q$ values and evaluated their RD performance separately. The results are depicted in Fig.~\ref{fig:heat}. It is evident that videos with lower spatio-temporal complexity tend to achieve better performance with smaller $f_q$ values, and vice versa. It also explains why the baseline with a fixed intensity, inevitably produces bad cases in videos with high spatio-temporal complexity. This observation aligns with our earlier analysis regarding processor intensity.
% \begin{figure}
%     \centering
%     \includegraphics[width=7cm]{siti_bdrate.png}
%     \caption{BD-Rate performance of baseline corresponding to SI/TI.}
%     \label{fig:siti-bdrate}
% \end{figure}
\vspace{-7pt}

\section{Conclusion}
\vspace{-6pt}

In this paper, we introduce a Tri-Dynamic Preprocessing framework for UGC video compression. The proposed framework can perceive spatio-temporal complexity in video sequences thus processing UGC data effectively with the help of the dynamic processing intensity, the dynamic training quantization level and the dynamic loss trade-off. Experimental results demonstrate that our proposed method consistently outperforms the anchor methods across a variety of commonly used perceptual metrics.

\clearpage

\vfill\pagebreak

% \section{REFERENCES}
% \label{sec:refs}

% List and number all bibliographical references at the end of the
% paper. The references can be numbered in alphabetic order or in
% order of appearance in the document. When referring to them in
% the text, type the corresponding reference number in square
% brackets as shown at the end of this sentence \cite{C2}. An
% additional final page (the fifth page, in most cases) is
% allowed, but must contain only references to the prior
% literature.

\bibliographystyle{IEEEbib}
\bibliography{strings,refs}

\end{document}